\documentclass[sw]{agutex}

\usepackage{rotating}
\usepackage{draftwatermark}
\usepackage{color, soul}
\definecolor{red}{rgb}{1,0,0}
\setulcolor{red}

\usepackage{graphicx}
\setkeys{Gin}{draft=false}
\authorrunninghead{BLAKE S.P. ET AL.}

\titlerunninghead{GICs in Detailed Irish Power Network Model}

\authoraddr{Corresponding author: Se\'{a}n P. Blake,
School of Physics, Trinity College Dublin, Dublin 2, Ireland
(blakese@tcd.ie)}

\begin{document}
%
%

\title{A Detailed Model of the Irish High Voltage Power Network for Simulating GICs}
%
%

%
%



 \authors{Se\'{a}n P. Blake,\altaffilmark{1,2}\thanks{Current address: NASA Goddard Space Flight Center, Space Weather Laboratory, Greenbelt, Maryland, USA}
 Peter T. Gallagher,\altaffilmark{1}\thanks{Current address: Astronomy and Astrophysics Section, School of Cosmic Physics, Dublin Institute for Advanced Studies, Dublin 2, Ireland},
 Joan Campany\`{a},\altaffilmark{1}$^{\dag}$
  Colin Hogg\altaffilmark{2},
 Ciar\'{a}n D. Beggan\altaffilmark{3},
 Alan W.P. Thomson\altaffilmark{3},
 Gemma S. Richardson\altaffilmark{3},
 David Bell\altaffilmark{4}}
 
\altaffiltext{1}{School of Physics, Trinity College Dublin, Dublin 2, Ireland}
\altaffiltext{2}{Dublin Institute for Advanced Studies, 5 Merrion Square, Dublin 2, Ireland}
\altaffiltext{3}{British Geological Survey, Lyell Centre, Riccarton, Edinburgh, EH14 4AP, UK}
\altaffiltext{4}{EirGrid Plc., The Oval, 160 Shelbourne Rd, Ballsbridge, Dublin 4, Ireland}





%
%


\keypoints{\item A detailed model of the Irish 400, 275, 220 and 110~kV power network was developed for GIC simulations.
\item The impact of assumptions used to construct a power network model were evaluated for GIC simulations.
\item Heavy rainfall before a 2015 geomagnetic storm may have decreased the grounding resistance at a substation, leading to larger GICs.}


%
%


\begin{abstract}

\noindent Constructing a power network model for geomagnetically induced current (GIC) calculations requires information on the DC resistances of elements within a network. This information is often not known, and power network models are simplified as a result, with assumptions used for network element resistances. Ireland's relatively small, isolated network presents an opportunity to model a complete power network in detail, using as much real-world information as possible. A complete model of the Irish 400, 275, 220 and 110 kV network was made for GIC calculations, with detailed information on the number, type and DC resistances of transformers. The measured grounding resistances at a number of substations were also included in the model, which represents a considerable improvement on previous models of the Irish power network for GIC calculations. Sensitivity tests were performed to show how calculated GIC amplitudes are affected by different aspects of the model. These tests investigated: (1) How the orientation of a uniform electric field affects GICs. (2) The effect of including/omitting lower-voltage elements of the power network. (3) How the substation grounding resistances assumptions affected GIC values. It was found that changing the grounding resistance value had a considerable effect on calculated GICs at some substations, and no discernible effect at others. Finally, five recent geomagnetic storm events were simulated in the network. It was found that heavy rainfall prior to the 26-28 August 2015 geomagnetic storm event may have had a measurable impact on measured GIC amplitudes at a 400/220 kV transformer ground. \textbf{Accepted for publication in AGU Space Weather. Copyright 2018 American Geophysical Union. DOI:10.1029/2018SW001926}

\end{abstract}

%
%

%

\begin{article}

%
%

\section{Introduction}

Geomagnetically induced currents (GICs) are one of the most disruptive and damaging space weather hazards. Variations in the Earth's magnetic field induce these electrical currents in grounded conductors such as railways \citep{Eroshenko2010}, pipelines \citep{Pulkkinen2001}, and particularly in power networks \citep{Pirjola2000}. GICs that arise during geomagnetic storm events can lead to transformer damage and widespread disruption to the network. The most famous example of the threat posed by GICs to power networks is the March 1989 geomagnetic storm, when GICs and their effects precipitated a blackout in the Hydro-Qu\'ebec transmission \citep{Bolduc2002}. 

The potential for damage to power networks has prompted studies of GICs around the world. It has long been recognised that the larger magnetic variations at higher latitudes drive larger GIC events, and studies have been been conducted in countries such as Finland \citep{Viljanen1994}, Sweden \citep{Wik2008}, Norway \citep{Myllys2014} and Canada \citep{Boteler1989}. It is now known that GICs can contribute to the failure of transformers in low-latitude and mid-latitude countries through repeated heating of transformer insulation \citep{Koen2003, Gaunt2007}. GICs can cause wear on transformers, leading to reduced efficiency and possible failure months after geomagnetic events, even if typical geomagnetic variations are small. GICs have been studied in power networks in lower latitude countries such as Austria \citep{Bailey2017}, Spain \citep{Torta2014, Torta2017}, China \citep{Zhang2015, Guo2015}, New Zealand, Australia \citep{Marshall2013}, South Africa \citep{Ngwira2011}, Ireland \citep{Blake2016}, the UK \citep{Beggan2013, Kelly2017}, and Brazil \citep{Barbosa2017}, among others.

The simplest way to study GICs in a network is to measure GICs as they flow to and from transformer grounds. This can be achieved using a Hall effect probe attached to a transformer ground. Ideally, every grounded point in a network would have a Hall effect probe for full GIC resolution. In practice, Hall effect probes can be expensive and disruptive to install on a transformer, and many countries have a limited number of Hall effect probes to measure GICs. A notable exception to this is New Zealand, which has tens of measurements which have been recording for several years \citep{MacManus2017}. GICs can also be measured by utilizing the differential magnetometer method \citep{Matandirotya2016}, where magnetometers measure the magnetic signal of GICs beneath transmission lines. Another indirect measure of GICs is to examine the chemical composition of gas in transformer housing \citep{Gaunt2014}. As the transformer is repeatedly heated from GICs, this chemical composition will change with time.

Where GIC measurements are limited to a few transformers (or none at all), GICs can be simulated in a network. These estimations are commonly separated into two distinct parts: the geophysical step, and the engineering step \citep{Pirjola2000}. The geophysical step involves calculating surface horizontal electric fields induced by the varying geomagnetic field. The magnitude of the induced surface electric fields depends on the resistivity of the subsurface geology \citep{Wei2013, Pulkkinen2012}. Different methods of calculating surface electric fields are used for GIC calculation, {including the simple plane-wave method \citep{Pirjola2001}, the multi-dimensional magnetotelluric (MT) method \citep{Bedrosian2015, Love2015, Torta2017} and the thin-sheet method \citep{Bailey2017, Thomson2005}.

Once the surface electric field is calculated, GICs in a grounded power network can be calculated (the engineering step). Accurately modelling a power network for GIC calculations requires knowledge of different components of the network. These include the transformer types in substations, the DC resistances of the windings in these transformers, the DC resistance of connections between substations, and substation grounding resistances \citep{Boteler2016}. Using this information, a model of a power network can be constructed. This can then be imposed upon surface electric field values, and GICs can be calculated for each grounded point.

Frequently, researchers do not have access to information on the elements of a power network for GIC modelling. As such, estimations of substation grounding and transformer winding resistances are often made for some or all of a power network \citep{Myllys2014, Torta2014, Blake2016}. In addition, studies often focus on the power network of a particular country in isolation, whereas in reality, most countries have interconnected power networks. The points of interconnection provide a route for GIC to flow between power networks, and so must be modelled for GIC calculations. Equivalent circuits can be constructed to approximate the connection between two different power networks \citep{Boteler2016}. Finally, when a power network is modelled, lower voltage regimes within the network are often omitted. High voltage transmission lines tend to be longer and have lower resistances than lower voltage lines, meaning they are more likely to experience the largest GIC values. Studies have shown that neglecting the lower voltage portion of a network can significantly change the calculated GIC in a network \citep{Torta2014, Guo2015}.

Ireland's relatively small network presents an opportunity to model a stand-alone power network while making minimal assumptions about the DC characteristics of its components.  Apart from two high voltage DC (HVDC) lines which connect it to Wales and Scotland, the integrated power network in Ireland and Northern Ireland is isolated from other power networks, and can be modelled without approximating adjacent power networks. In this paper, we present a detailed model of the Irish power grid for GIC calculations. This improved model includes 400, 275, 220 and 110~kV lines and substations, and models the number and type of transformers in substations that operate at 220~kV or higher. Information on the grounding resistances was provided by EirGrid (Ireland's power network operator) for a limited number of substations, and Tee-junction connections were used to more accurately model GICs. A number of different sensitivity tests were performed on the power network model to investigate how the assumptions used in its construction affect calculated GIC values. Finally, GICs were calculated in the network for five minor geomagnetic events, and the values were compared to GIC values measured at a single transformer in Ireland.

\section{Modelling GICs in a Power Network} \label{Modelling_Section}

\begin{center}
\begin{figure}
\noindent\includegraphics[width=22pc]{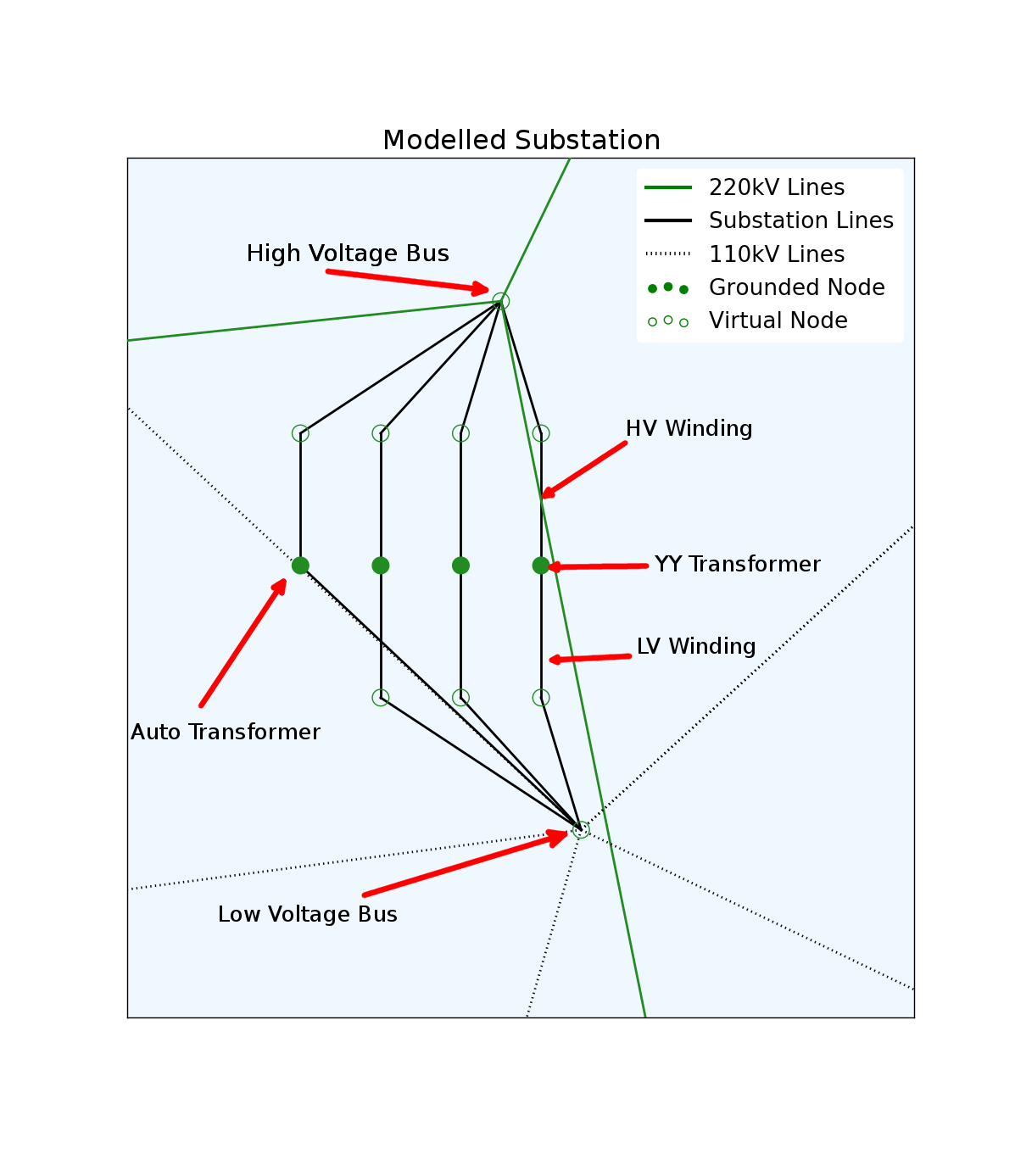}
\caption{A 220/110~kV substation created by the power network model generator. This substation has one auto-transformer, and three YY-transformers connected in parallel. The autotransformer has one internal connection which has the resistance of the HV winding. Each of the YY-transformers has two connections with resistances set to the HV and LV winding resistances. The resistances of the substation lines between buses and nodes are set to be infinitesimally small.}
\label{PNMG}
\end{figure}
\end{center}

While power network operators often use commercially available power system analysis tools to analyse GICs, the most commonly used approach to modelling GICs in academia is the Lehtinen-Pirjola (LP) method \citep{Lehtinen1985}. This approach treats a subject power grid as a discretely earthed network, and applies Ohm and Kirchoff's Laws in order to calculate induced currents. As GICs are driven by magnetic field variations with frequencies $<$1~Hz, it is appropriate to treat GICs as DC \citep{Boteler2016}. The LP method allows for the modelling of any power network, so long as the following information about the network is known (or estimated): positions of substations, types and numbers of transformers in substations, transformer winding resistances, substation grounding resistances, connections between substations and the resistances of these connections.

This information can be used to solve the following for GICs in a network:

\begin{equation}
 \mathbf{I} = (\mathbf{1} + \mathbf{YZ})^{-1}\mathbf{J}
\end{equation}

where $\mathbf{I}$ is the matrix of GIC values flowing through earthed nodes (transformers), $\mathbf{1}$ is the unit matrix, $\mathbf{Y}$ is the network admittance matrix (defined by the resistances of the conductors of the network), $\mathbf{Z}$ is the earthing impedance matrix and $\mathbf{J}$ is the `perfect earthing' current, defined as:

\begin{equation}
 J_{i} = \sum_{j \neq i} \frac{V_{ij}}{R_{ij}}
\end{equation}

where $V_{ij}$ and $R_{ij}$ refer to the geo-voltages and line resistances between two nodes $i$ and $j$ \citep{Beggan2015}.  AC power networks utilize three phase power lines. In order to simplify the calculation of GICs in these lines, the parallel paths of each phase can be used to calculate an equivalent circuit for GIC calculation \citep{Boteler2016}. In practical terms, this involves dividing line and transformer winding resistances by three. In most transmission networks, there are different voltage levels which can be modelled. The flow of GICs in a network of multiple voltages will be through the windings of the transformers at each substation. The type of transformer determines that nature of the path for flow of GICs. \citet{Boteler2014} outlines how two-winding (or `YY') and auto-transformers can be treated in a power network model in order to more accurately simulate GICs with the LP method.}

By including virtual nodes (nodes with infinitely large grounding resistances) at the neutral points of transformers, YY and auto-transformers can be modelled without introducing non-zero off-diagonal elements in the earthing impedance matrix $\mathbf{Z}$ \citep{Pirjola2005}. For auto-transformers, a single virtual node is placed at the high-voltage connection point of the transformer. For a YY-transformer, a node is placed at both the high and low-voltage connection points of the transformer.

There are two challenging aspects to modelling a power network for GIC calculations when utilizing the LP method. The first is collating the information regarding the power network. As power networks typically have hundreds of substations, this can be a time-consuming task to gather DC characteristics of a power network (particularly if this information is not digitised). Assuming one can collect this information, the second challenging task is constructing the network model so that it can be used to calculate GICs with the LP method in the manner outlined above. This process was automated using an open-source Python program (https://doi.org/10.5281/zenodo.1252432). This program takes as inputs the collated information on a power network, and outputs a model that can be used with the LP method of GIC calculation. The program can take into account the different transformer types listed above, multiple transformers per substation and dual-circuit connections between substations. An example of how the program handles a substation with multiple transformers is shown in Figure~\ref{PNMG}. This shows a 220/110~kV substation with four transformers connected in parallel (one auto and three YY-transformers). They are each connected to high and low voltage buses, which connect to other 220 and 110~kV substations respectively. The filled circles are connected to a common ground, and it is here that the GIC are calculated. Each of the unfilled circles are the virtual nodes with infinite grounding resistances. The connections between nodes were given resistances equal to the winding resistances of the transformers.

\begin{center}
\begin{figure}
\noindent\includegraphics[width=20pc]{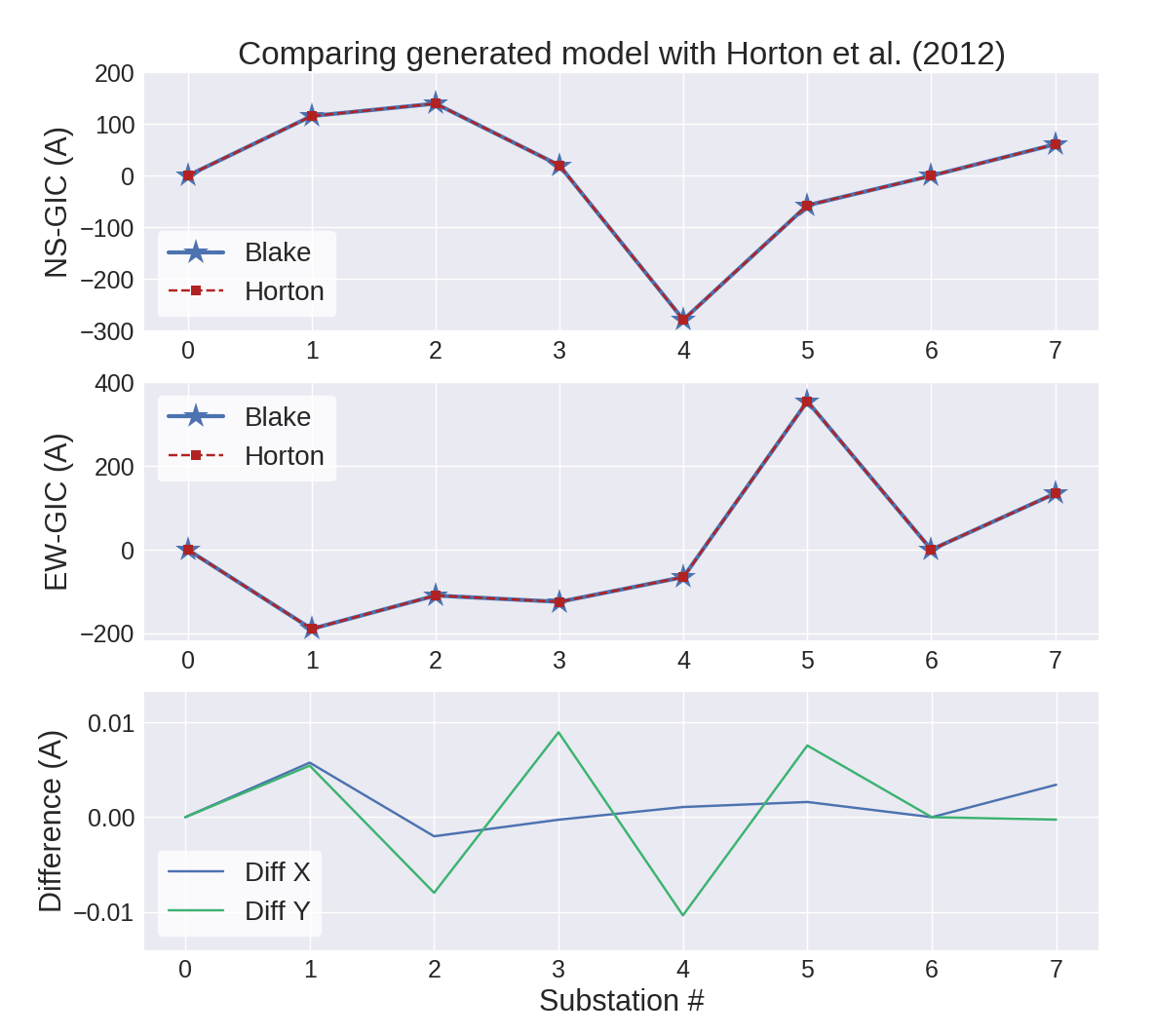}
\caption{GIC values given in \citet{Horton2012} compared to GIC values calculated using the power network model generator.}
\label{Horton_confirmation}
\end{figure}
\end{center}

In order to verify that the program was accurately creating power network models for given transformer and connection information, the test-case 500 and 345~kV power network featured in \citet{Horton2012} was used as an input to the network model generator program. This network consists of eight substations with a total of 15 auto and YY-transformers. Also included in this model is a GIC blocking device at one substation, and some instances of multiple parallel connections between substations. This information was fed into the model generator, and the output model was subjected to a 1~V~km$^{-1}$ electric field. The resulting GICs calculated at each of the substations were compared to those values calculated in \citet{Horton2012}. Figure~\ref{Horton_confirmation} shows the comparison between the GIC values given in the paper and those calculated using the power network model generator. The differences between the calculated GIC and the \citet{Horton2012} model are on the order of 0.01~A. As the differences in calculated GIC values were small given the amplitude of modelled GICs, the power network model generator was used to create a model of the Irish power network using all available information provided by EirGrid.

%

\section{Constructing a Detailed Model of The Irish Power Network} \label{CONSTRUCTING_SECTION}

\begin{center}
\begin{figure}
\noindent\includegraphics[width=20pc]{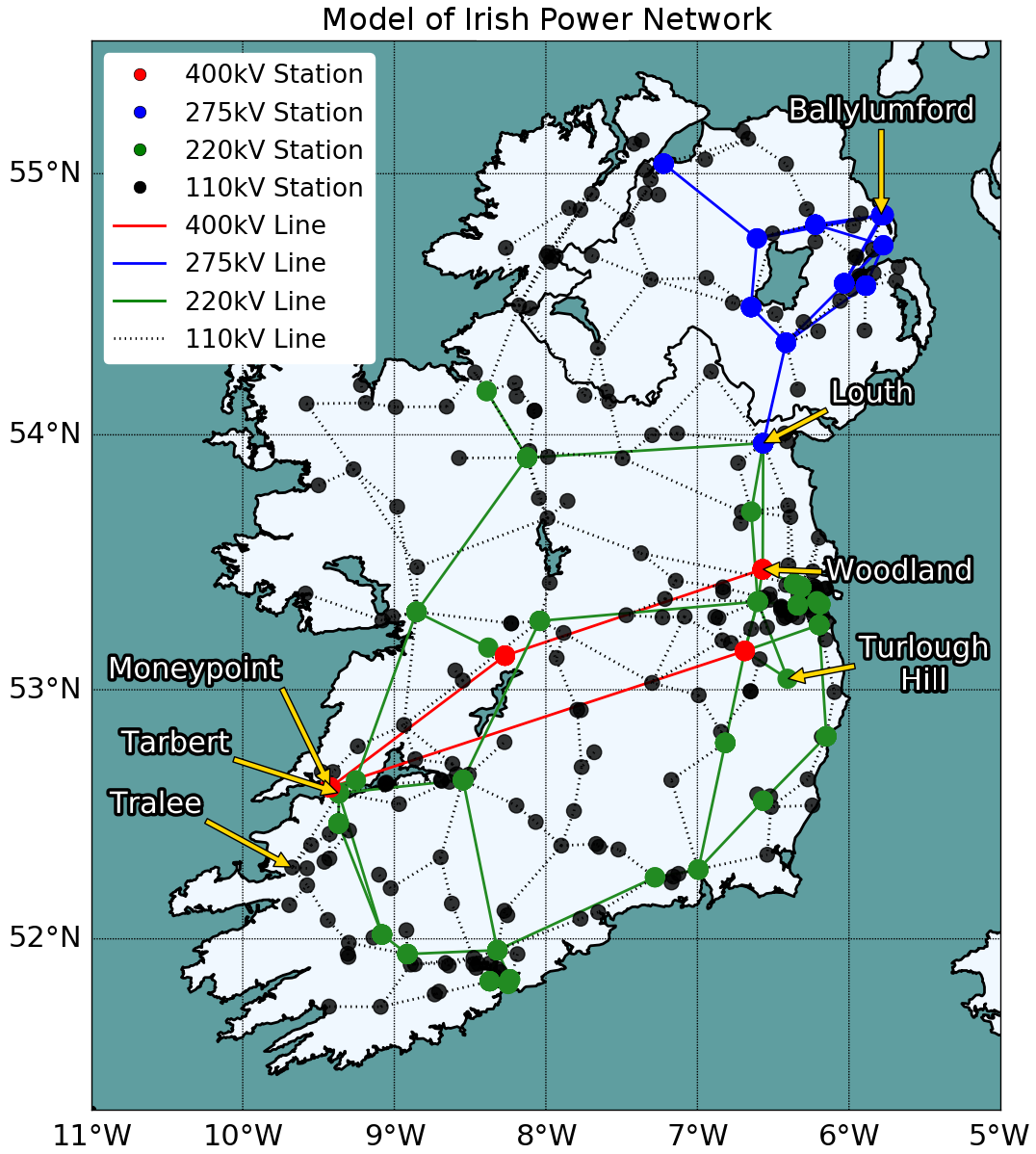}
\caption{Model of the Irish HV power network. This model includes 400, 275, 220 and 110~kV substations and lines. Ireland's only Hall Effect GIC monitor is located at the 400/220~kV Woodland substation.}
\label{Irish_model}
\end{figure}
\end{center}

The Irish power network consists of approximately 270 substations and 6,400~km of 400, 275, 220 and 110~kV transmission lines in both countries of Ireland and Northern Ireland. The all-island power network is isolated from other power networks (except via two HVDC connections), so it can therefore be considered as a whole without approximating peripheral power networks. Ireland's small size (approximately 500 $\times$ 300~km) and population means that it requires fewer high-voltage lines and substations than other larger countries. As such, Ireland has only four 400~kV substations, and three 400~kV transmission lines, running roughly West-East. The next highest voltage lines (275~kV) operate only in Northern Ireland, and 220~kV substations operate only in the Republic of Ireland. 110~kV substations and lines operate on all parts of the island of Ireland, and in low population density areas (such as the North West and West), 110~kV lines are the only high voltage (HV) transmission lines in operation. The power network can be seen in Figure~\ref{Irish_model}.

The Irish power network model used in \citet{Blake2016} did not include 110~kV substations and transmission lines. In addition, it assumed a single transformer per substation, as well as resistance values for both transformer windings and substation grounds (0.5~$\Omega$ and 0.1~$\Omega$ respectively). This model has been improved upon to include the following: (1) The correct number and type of transformers in substations which operated at 220~kV or higher. (2) The DC resistances of the high and low voltage windings in these transformers. These values ranged from 0.04 to 0.68~$\Omega$. (3) The substation grounding resistance measured at 33 substations across Ireland. These values ranged from 0.25 to 6.35~$\Omega$. The remaining 237 substations were given grounding resistances fixed at 1~$\Omega$. This value was chosen as it is the value that EirGrid aim to maintain at their substations for operational safety. It is worth noting that these values are considerably larger than the 0.1~$\Omega$ typically used in studies when true grounding resistances are unknown. (4) DC resistances for all transmission lines (including instances where multiple connections exist between substations). (5) A number of T-junctions in the network (modelled as nodes with infinitely large grounding resistances).

Each substation that operates at 110~kV was assumed to have a single transformer with a winding resistance of 0.087~$\Omega$. This value was chosen as it was a representative resistance value of the LV windings in the 220/110~kV transformers. All of this information was used as inputs for the power network model generator, and the output model was used for GIC calculations. Of all of the substations in the network, Louth is the only substation with transformers operating at three different voltages(275, 220 and 110~kV). This was therefore modelled as two separate substations (275/220 and 220/110~kV) with a shared grounding resistance value. For the purpose of analysis, GICs calculated at the two substations were summed.

\section{Sensitivity Tests} \label{SENSITIVITY_TEST_SECTION}

Once a network model has been created, it is informative to subject it to idealised geoelectric fields and subsequently calculate GIC values at the model nodes. This will give an indication as to which substations will see larger GIC amplitudes due solely to the orientation of a network. This exercise has been used on a number of different network models in different studies \citep{Horton2012, Myllys2014, Torta2014, Blake2016}.

\subsection{Electric Field Orientation}
Uniform 1 V~km$^{-1}$ electric fields oriented North and East were applied to the model network, and resulting GICs were calculated at each of the nodes. The calculated GICs for each substation can be found in Figure~\ref{1VKM}. The Moneypoint substation (substation number 2 in figures below) had the largest calculated GICs, with a maximum of 114~A for an eastward directed electric field. The large GIC values in this substation are to be expected, given that it is connected to two of the largest transmission lines in the country. The substation that experienced the next largest GIC was the 275/220~kV Louth substation, which connects the HV networks in the Republic and Northern Ireland. GICs were calculated at 38~A when the uniform electric field was oriented northward. Only a single 110~kV substation had calculated GIC values greater than 25~A. This was Tralee in the South-West, with calculated GICs of 28~A for an eastward directed field.

\begin{center}
\begin{figure}
\noindent\includegraphics[width=20pc]{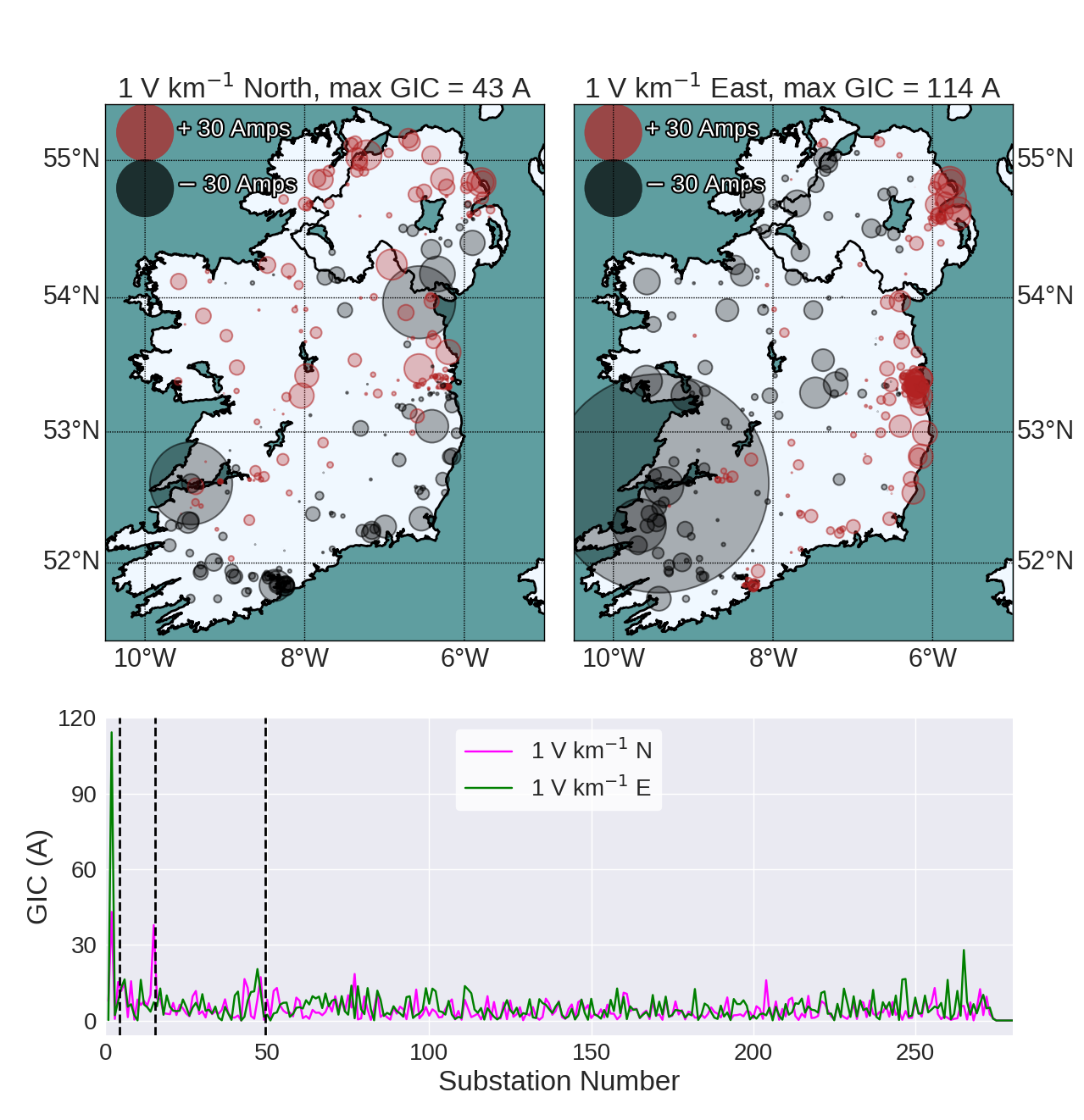}
\caption{Response of the Irish power network model to uniform 1~V~km$^{-1}$ electric fields pointing North and East. The model saw peak GICs of 113~A in the 400/220~kV Moneypoint substation in the West of Ireland. The bottom plot shows the calculated GIC for each substation. The dashed lines separate (from left to right) the 400, 275, 220 and 110~kV substations. These are ordered alphabetically within each voltage division.}
\label{1VKM}
\end{figure}
\end{center}

\begin{center}
\begin{figure}
\noindent\includegraphics[width=20pc]{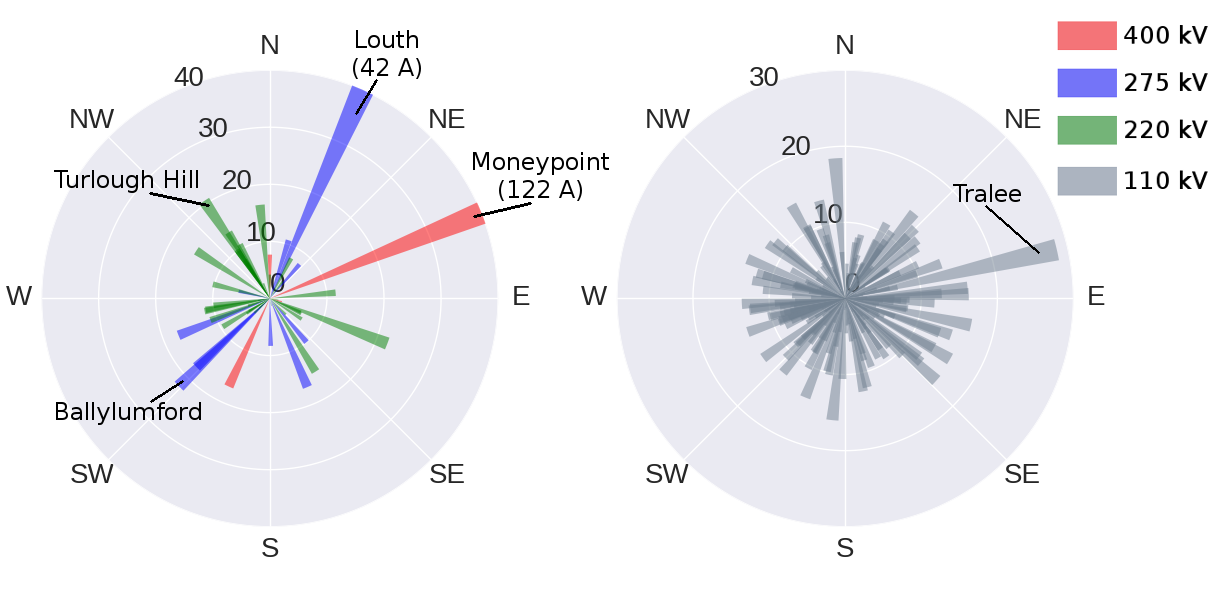}
\caption{Maximum positive GIC calculated for each substation when a uniform 1~V~km$^{-1}$ electric field is rotated 360$^{\circ}$ from North. In each subfigure, angle corresponds to direction of electric field, and length of each segment is the maximum current. The red, blue, green and grey segments correspond to the 400, 275, 220 and 110~kV substations respectively. GICs of 122~A were calculated at the Moneypoint substation.}
\label{ROSE}
\end{figure}
\end{center}

Figure~\ref{ROSE} shows the maximum positive calculated GIC value at each substation when the 1~V~km$^{-1}$ is rotated 360$^{\circ}$. Moneypoint again had the largest calculated GIC values, with 122~A when the electric field points 69$^{\circ}$ clockwise from North. With the exception of Tralee, it can be seen that the 275 and 220~kV substations experience marginally larger GICs than the 110~kV substations. With the exception of Moneypoint, Ireland has relatively low calculated GIC values for a 1~V~km$^{-1}$ electric field when compared to other countries \citep{Myllys2014, Torta2014}. This is likely due to a combination of factors. The network's small size limits the length of the largest transmission lines in Ireland to less than 200~km. With the addition of the 110~kV substations, Ireland has approximately one grounded transformer for every 220~km$^{2}$, providing many grounded points in the network for GICs to flow to and from the ground. Additionally, as noted above, the grounding resistance values used in the model are larger than those used in other studies. This will also limit calculated GIC amplitudes.

\subsection{Including Lower Voltage Elements}
As shown in \citet{Guo2015} and \citet{Torta2014}, neglecting lower-voltage elements in a power network can have a significant impact on calculated GICs, although this is highly dependent on the makeup of an individual network. In order to investigate the effect including lower voltage elements have on GIC calculations in the Irish power network, three calculations were made for northward and eastward-directed uniform electric fields. In the first calculation, only the 400~kV substations and lines were included in the power network. After this, the model was altered to include the 275 and 220~kV elements. Finally, the 110~kV substations and lines were included to complete the HV network model.

\begin{center}
\begin{figure}
\noindent\includegraphics[width=22pc]{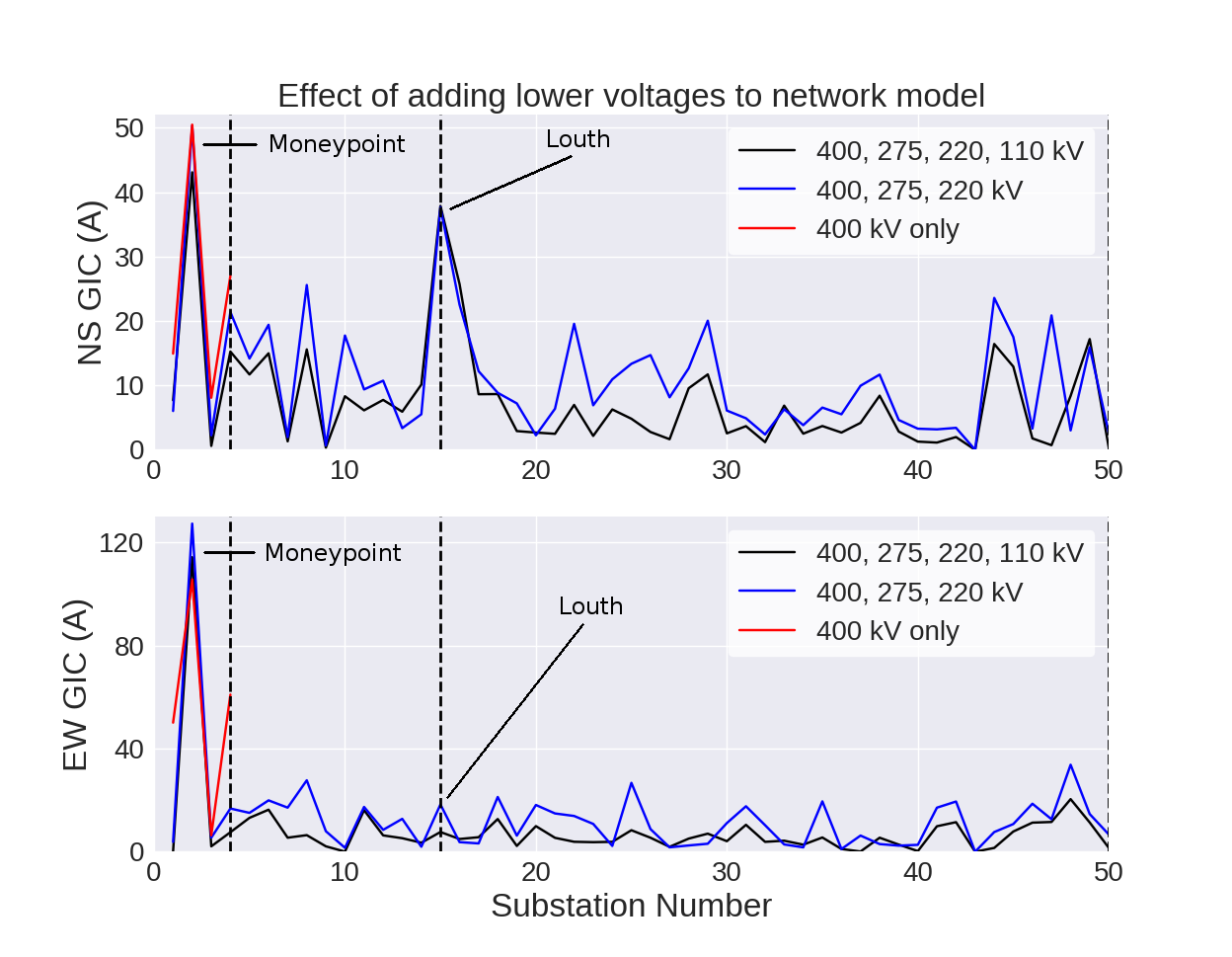}
\caption{Calculated GICs for different voltage levels in the Irish network. The top and bottom plots are for northward and eastward electric fields respectively. The dashed lines separate (from left to right) the 400, 275 and 220~kV substations.}
\label{VOLTAGES}
\end{figure}
\end{center}

Figure~\ref{VOLTAGES} shows the calculated GICs for each of these three network models when a uniform 1~V~km$^{-1}$ electric field is imposed on the network. When the 400~kV only model is compared to the 400, 275 and 220~kV model, each of the the 400~kV substations have smaller calculated GICs. The exception to this is the Moneypoint substation when the uniform electric field is eastward-oriented. In this case, GICs at Moneypoint increased from 105~A to 127~A with the addition of the 275 and 220~kV elements of the network. Generally, for each of the 400, 275 and 220~kV networks, the addition of the 110~kV network decreases the calculated GICs, as the currents are directed to and from the lower voltage subsations. The ten substations which had the largest change in GICs amplitudes each saw reduced GICs. Each of these ten substations were connected to at least three 110~kV substations (an exception to this is Moneypoint, which is connected only to a single 220~kV substation). It is to be expected that substations most affected by the addition of the lower voltages are connected to lower voltage substations.

\begin{center}
\begin{figure}
\noindent\includegraphics[width=20pc]{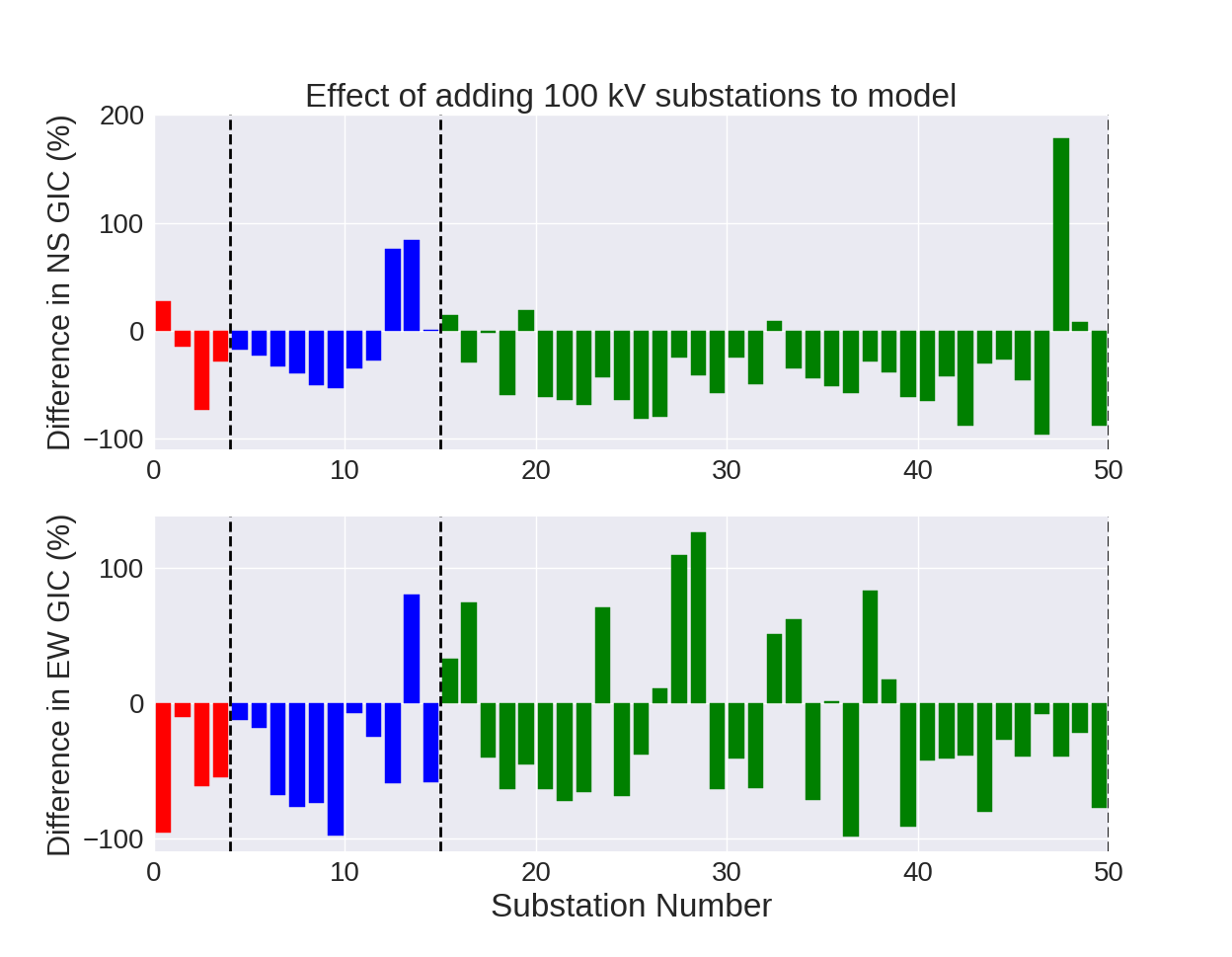}
\caption{Proportional difference in calculated GICs when the 110~kV elements are added to the Irish power network model. The top and bottom plots are for northward and eastward directed electric fields respectively. The dashed lines separate (from left to right) the 400, 275 and 220~kV substations.}
\label{VOLTAGES_PROP}
\end{figure}
\end{center}

The proportional changes in calculated GICs when the 110~kV substations are added are shown in Figure~\ref{VOLTAGES_PROP}. This shows that there was a general decrease in GIC although there were increased GIC amplitudes in 17 substations. The largest proportional increase occurs in the 220~kV Tarbert substation, increasing by 178\% with a northward-directed field. Whilst this is quite a large proportional increase, in absolute terms, the GICs at this substation increased from 3.01 to 8.36~A. In the case of the Irish power network, with some exceptions, omitting the lower voltage elements in the network model would lead to overestimating the `true' GIC values in higher voltage substations, particularly at those substations with direct connections to the lower voltage elements.

\subsection{Grounding Resistances} \label{GROUND_SECTION_REF}
\begin{center}
\begin{figure}
\noindent\includegraphics[width=20pc]{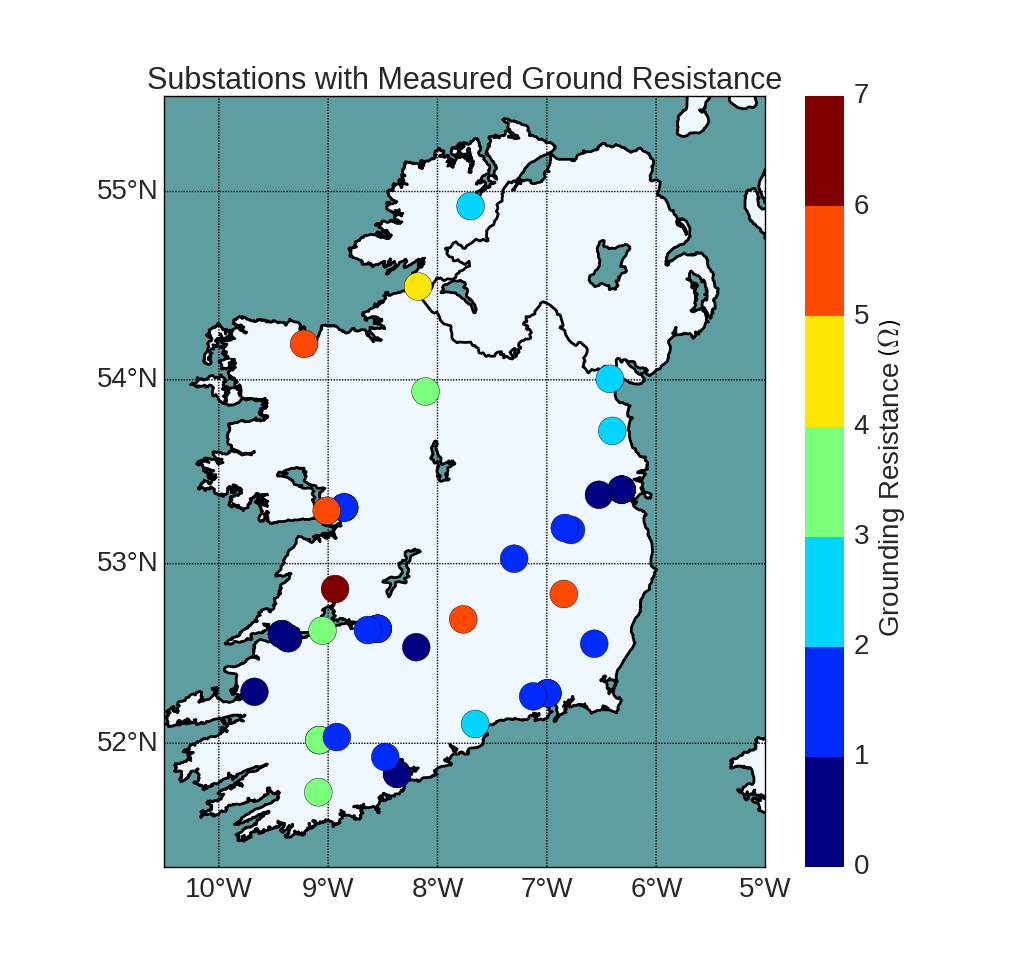}
\caption{Locations of substations with measured grounding resistances. These values range from 0.25--6.35~$\Omega$. The remaining substations were set at 1~$\Omega$.}
\label{GROUNDS}
\end{figure}
\end{center}

\begin{center}
\begin{figure*}
\noindent\includegraphics[width=40pc]{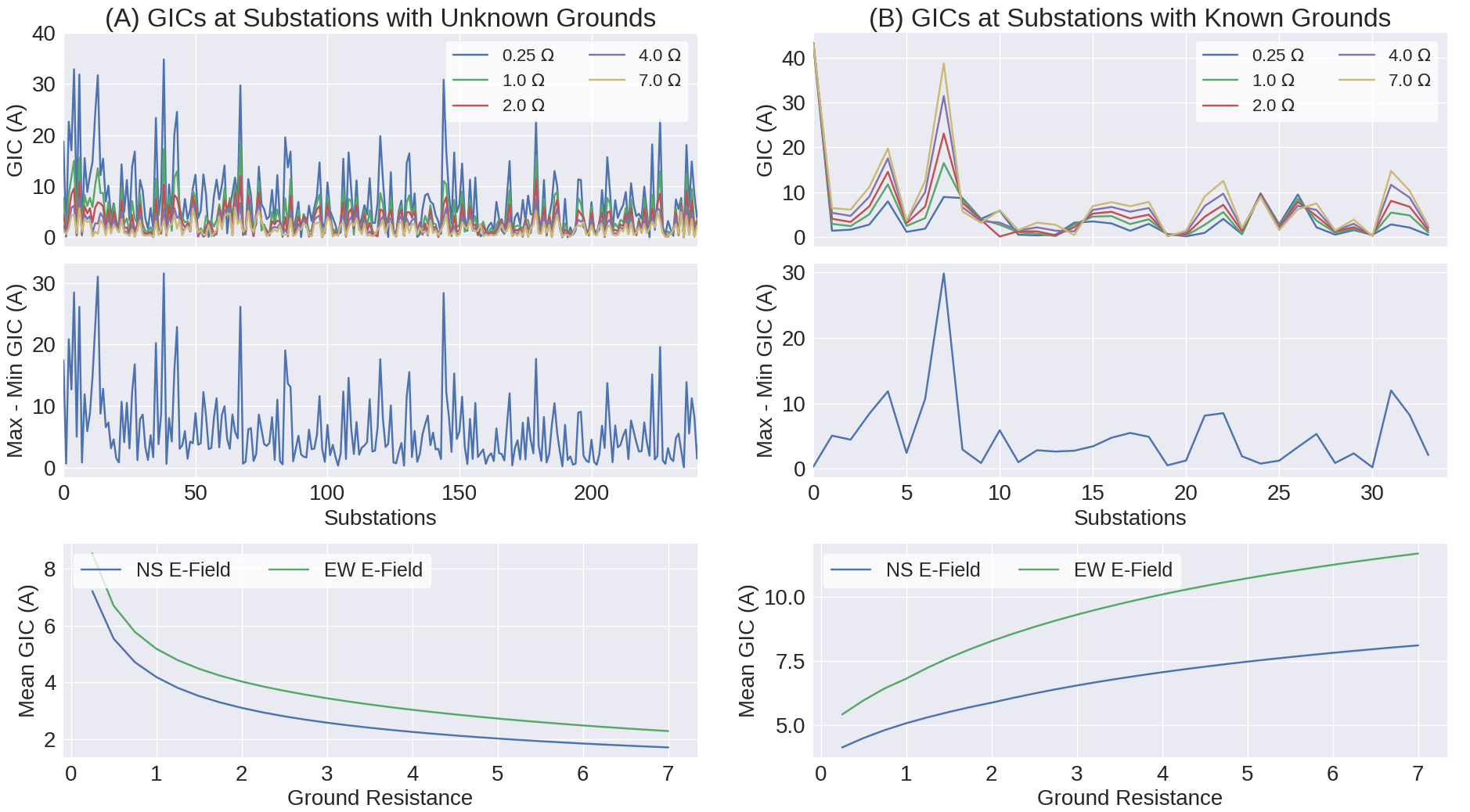}
\caption{The GIC response of substations due to a uniform 1~V~km$^{-1}$ electric field when the grounding resistance assumption of 1~$\Omega$ is altered. Subfigure (A) shows the response of those substations whose grounding resistances are unknown, subfigure (B) shows the response of the 33 substations with known and fixed grounding resistances. The bottom plots show the average GIC per substation as the grounding resistance assumption is altered.}
\label{GROUND_GIC}
\end{figure*}
\end{center}

The locations of the 33 substations which have known grounding resistances are shown in Figure~\ref{GROUNDS}. All other substations in the power network model were assumed to have a grounding resistance of 1~$\Omega$. When compared to the mean three-phase resistances for the transmission lines and individual transformer windings in the network (0.73~$\Omega$ and  0.062~$\Omega$ respectively), this large grounding resistance assumption will be an important factor in determining the distribution of GICs in the power network. In order to investigate how the assumed grounding resistance value affects GIC calculations, the 1~$\Omega$ assumption was varied from 0.25 to 7~$\Omega$ in 0.25~$\Omega$ increments. For each grounding resistance value chosen, a uniform 1~V~km$^{-1}$ electric field was applied to the network, and resulting GICs were calculated. Figure~\ref{GROUND_GIC} shows the calculated GICs at the substations with assumed and varying (subfigure A) and known and fixed (subfigure B) grounding resistances. 

Figure~\ref{GROUND_GIC} shows from top to bottom: the GIC calculated at each substation for a varying grounding resistance value, the variation in GIC for each substation (maximum calculated GIC minus minimum calculated GIC) for the simulations and assumed grounding resistance value against average GIC per substation. As can be expected, varying the grounding resistance assumption from 0.25~$\Omega$ up to 7~$\Omega$ gives greatly different GIC values at many of the substations. This is not true for all substations however, with GIC amplitudes varying by only a small amount at some substations. The Moneypoint substation is an example of one of the substations with a known grounding resistance value of 0.25~$\Omega$. Varying the grounding resistance of the other substations in the network changed the calculated GIC in Moneypoint by only 0.7~A. 

The variations in GIC amplitudes ranged from approximately 0 to 30~A for both groups of substations. This shows that the grounding resistance at a substation can be an important factor in GIC calculation, but the impact it has depends on the substation in question, and its connections to other substations. Generally, the larger the grounding resistance assumption, the smaller the GICs in the substations with unknown grounds. The inverse of this is true at the 33 substations with fixed grounds. By assuming large grounding resistances for the majority of the network, those substations with fixed values become more viable routes for GICs to flow to and from the ground.

\begin{center}
\begin{figure*}
\noindent\includegraphics[width=40pc]{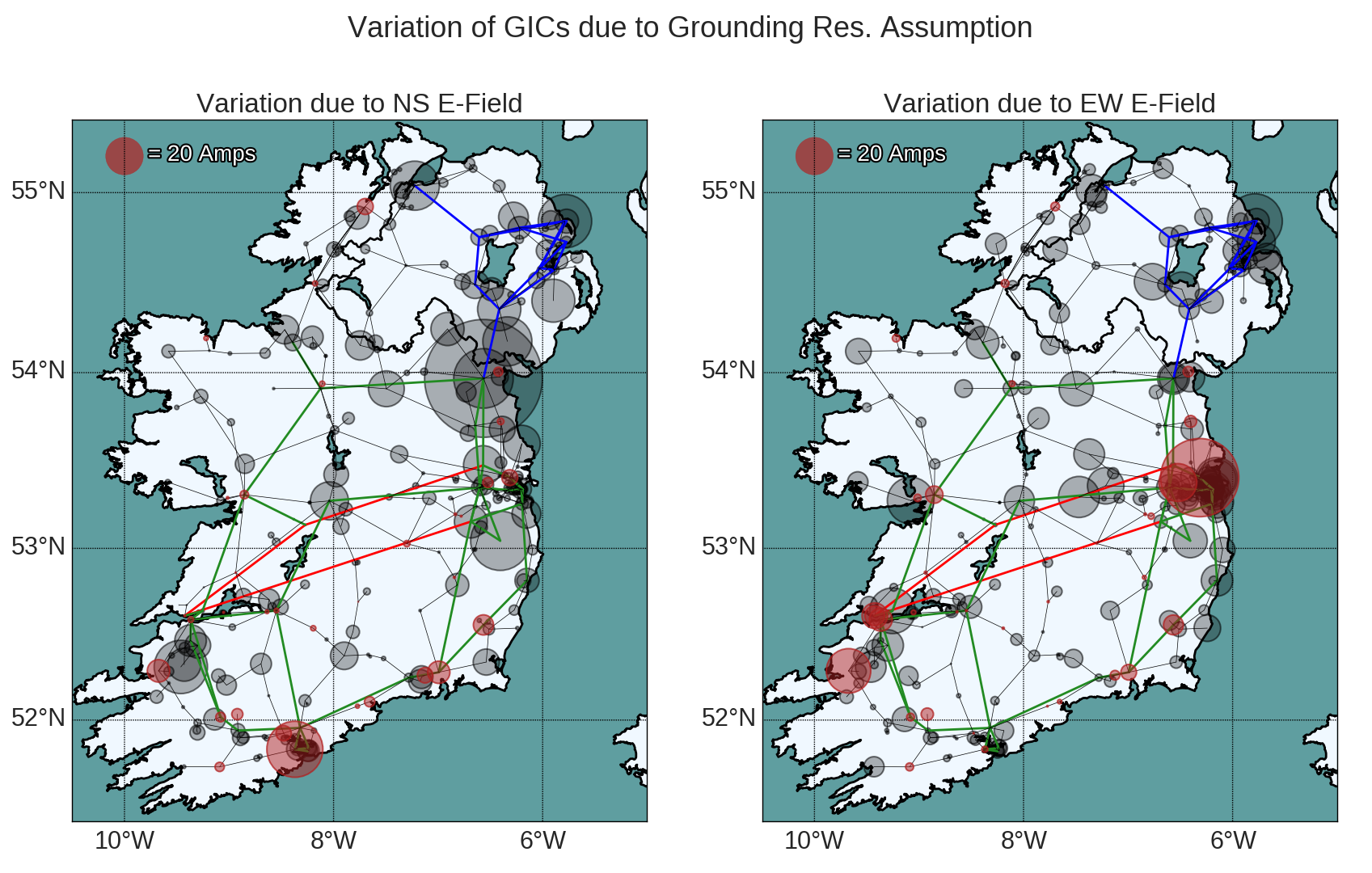}
\caption{Maximum variations in calculated GICs for different grounding resistance assumption. The grounding resistances were varied from 0.25--7~$\Omega$. Red circles are for substations with known grounds, and black are for those with assumed grounding resistances.}
\label{GROUND_GIC_SPATIAL}
\end{figure*}
\end{center}

Figure~\ref{GROUND_GIC_SPATIAL} shows the the maximum calculated variation for each substation for northward and eastward directed electric fields. The red circles indicate substations with known, fixed grounding resistances, and the black circles indicate those substations whose grounding resistance was varied. Figure~\ref{GROUND_GIC_SPATIAL} shows that the orientation of the electric field will also determine how much of an effect varying the grounding resistance will have on calculated GICs: some substations show large variations with northerly electric fields, but not with easterly electric fields, and vice versa.

\section{Modelling GICs During Recent Geomagnetic Storms}
In mid-2015, a Hall Effect probe was installed on a transformer ground in the 400/220~kV Woodland substation in the East of Ireland. Since then, the device has been recording continuously, and has been operational for a number of minor geomagnetic storm events. Five of these events were used to simulate GICs in the complete HV power network described above. These events are the 26-28 August 2015, 07-08 September 2015, 07-08 October 2015, 20-21 December 2015, and the 06-07 March 2016 storms. The planetary K (Kp) and disturbed storm time (Dst) indices for each event are given in Table~\ref{GOF}.

\begin{center}
\begin{figure*}
\noindent\includegraphics[width=30pc]{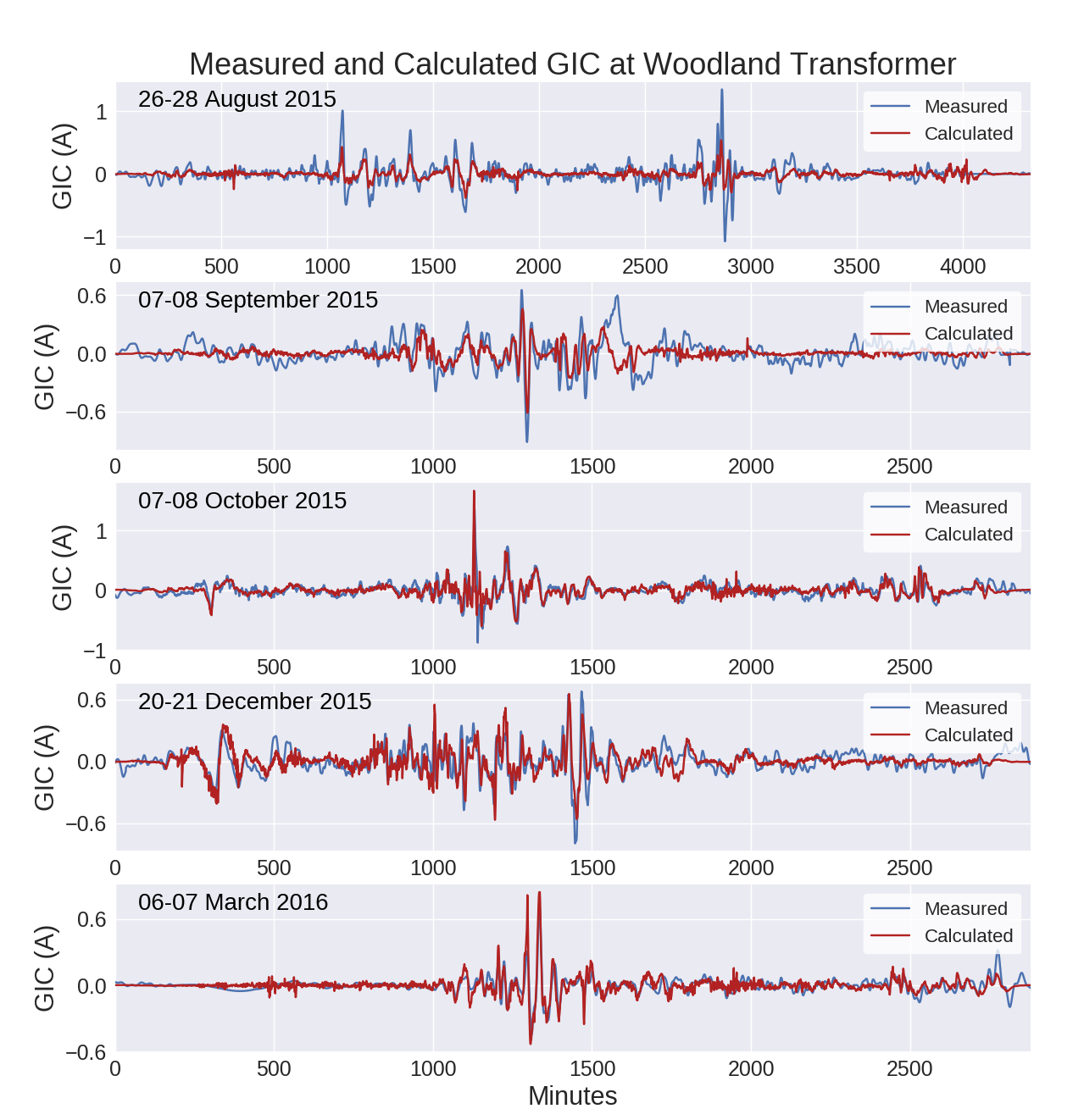}
\caption{Measured and calculated GIC at a transformer in the 400/220~kV Woodland substation for five recent storm events. The GICs were driven by electric fields calculated using a uniform 400~$\Omega$m resistivity model.}
\label{EVENTS}
\end{figure*}
\end{center}

For each event, magnetic data were collected from the INTERMAGNET and MagIE \citep{Blake2016} observatory networks around Ireland, Britain and Europe. These data were interpolated across Ireland using the spherical elementary current system method \citep{Amm1999}. A uniform 400~$\Omega$m ground resistivity model was used with the magnetotelluric method to calculate surface electric fields across Ireland. Finally, the model of the complete power network was overlaid on the calculated surface electric fields, and GICs were calculated. The calculated values at the Woodland transformer were then compared to the measured GIC values. The measured and calculated GIC values at the Woodland transformer for each of the five events are shown in Figure~\ref{EVENTS}.

Three goodness-of-fit measures were selected to quantify the fit of the calculated GIC to the measured GICs. These are the root mean squared error ($\textrm{RMSD}$), the \citet{Torta2014} defined performance parameter ($P$), and Pearson's correlation coefficient ($R$). The goodness-of-fit measures for each of the events are shown in Table~\ref{GOF}. Generally, the measured GICs during each of the storm events were modelled reasonably well using the 400~$\Omega$m resistivity model and the detailed power network model, although different events are modelled to a different degree of accuracy. Of the five events, the worst-fit was the 07-08 September 2015 event, with the highest RMSD and lowest $R$ value. The other four events had correlation coefficients which ranged from 0.59 to 0.68.

\begin{center}
\begin{table*} 
\centering
\caption{Goodness of fit measures for the calculated and measured GICs at Woodland for five recent geomagnetic storm events. These are the root-mean-square-difference (RMSD),  the \citet{Torta2014} defined performance parameter ($P$), and Pearson's correlation coefficient ($R$), where subscripts $o$ and $c$ refer to the observed and calculated GIC values. Kp and Dst values for each storm were taken from http://wdc.kugi.kyoto-u.ac.jp/.}
\label{GOF}
 \begin{tabular}{|c||c|c|c|c|c|}
 \hline
 \textbf{Storm Event}& Kp & Dst (nT) & RMSD$_{oc}$ (A)   & $P_{oc}$    & $R_{oc}$\\ \hline
\textbf{26-28 Aug. 2015} & 6+  & $-90$ & 0.127    & 0.224    & 0.685 \\
\textbf{07-08 Sep. 2015} & 6+  & $-70$ & 0.128    & 0.090    & 0.425 \\
\textbf{07-08 Oct. 2015} & 7+  & $-124$ & 0.100    & 0.239    & 0.676\\
\textbf{20-21 Dec. 2015} & 7$-$& $-155$ & 0.101    & 0.201    & 0.639\\
\textbf{06-07 Mar. 2016} & 6+  & $-98$  & 0.071   & 0.044    & 0.599\\ \hline
\end{tabular}
\end{table*}
\end{center}

While the detailed power network model was able to replicate GICs with a reasonable level of accuracy for five small events, there are a number of caveats which must be taken into account. Firstly, Ireland has only a single GIC probe at a single transformer. Uncertainty exists around the calculated GIC values in the rest of the power network. Secondly, the GICs seen in Woodland since 2015 are all quite small. No GIC values larger than 2~A have yet been measured in Ireland. It is worth noting that the low measured GIC amplitudes in the five events are approximately an order of magnitude larger than the noise levels in the Hall Effect probe ($\pm$0.1~A). The accuracy of the model for larger or historical events (pre-2015) can only be estimated. Finally, the five geomagnetic events in this paper occur over a period of eight months. During this time, it is possible that the network may have had elements added or removed as part of its normal operation (either for repairs/maintenance or load balancing), altering the flow of GICs in the network. This information was not available for modelling however, and the same network model was used for all five events. This may explain why some events are better modelled than others. 

\subsection{Soil Moisture Variation and GIC Amplitudes}
The August 2015 event is of particular interest, as unlike the other four simulated events, it has a high correlation coefficient (0.68), but it underestimates the amplitude of GICs by a factor of approximately 2.5. The difference between this event and the other four may be due to the variability of Woodland's ground resistance due to terrestrial weather. If the sediment beneath a substation becomes sodden due to rainfall, it will become more conductive. This will reduce the grounding resistance between the transformer neutrals and the ground, allowing larger GICs to enter the network. In the five simulations, the Woodland substation was given a grounding resistance of 0.5~$\Omega$. In reality, this may have differed for each event, depending on the conductivity of the subsurface at the time of the storm.

\begin{center}
\begin{figure*}
\noindent\includegraphics[width=30pc]{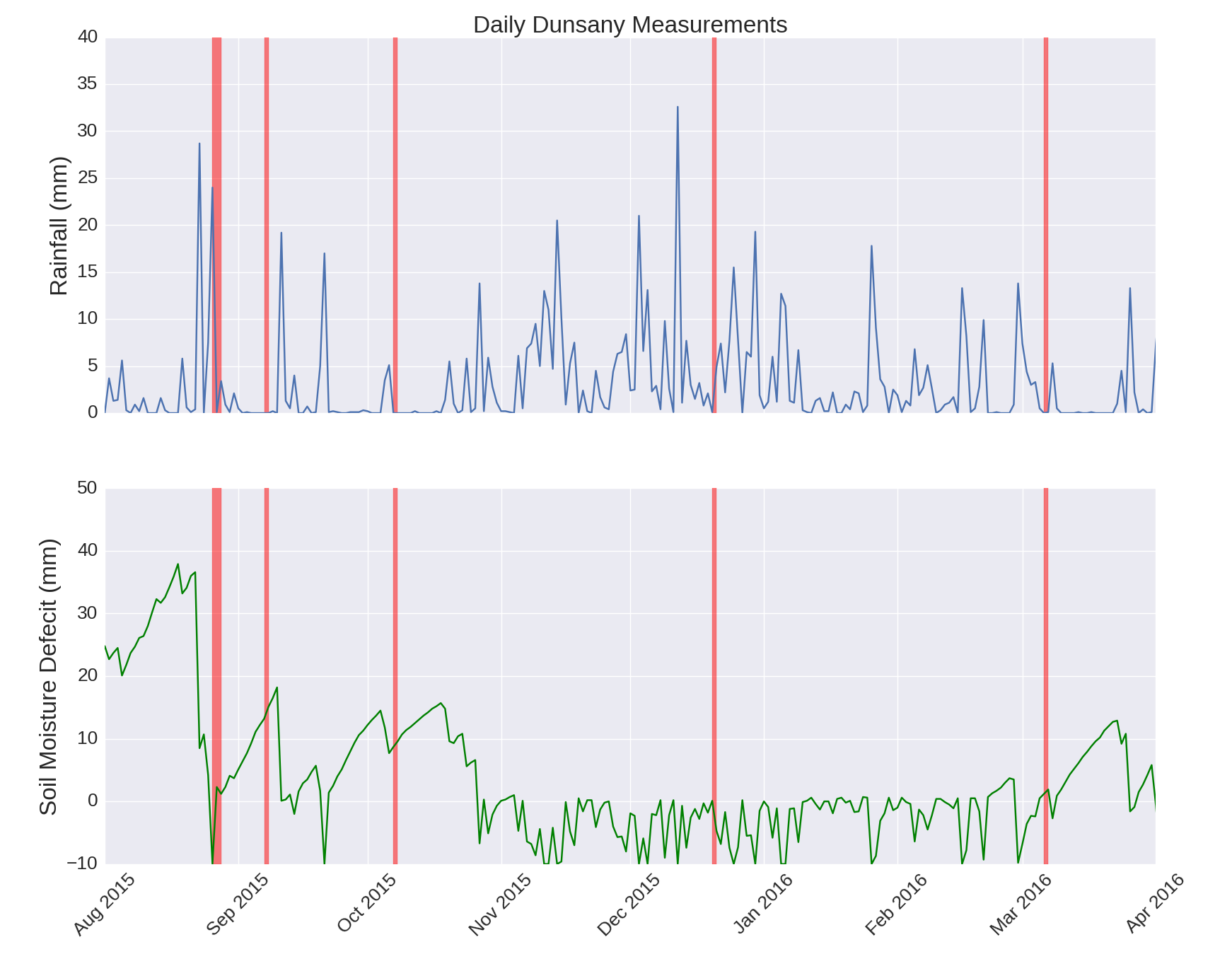}
\caption{Rainfall (top) and soil moisture deficit for a moderately drained soil sample (bottom) at the Met \'{E}ireann operated meteorological station 7~km from the Woodland substation. The times highlighted in red are the five geomagnetic storm events modelled in this paper. 60~mm of rain fell in the three days before the August 2015 geomagnetic storm event, and the soil sample at Dunsany was saturated for the first day of the geomagnetic storm event.}
\label{RAINFALL}
\end{figure*}
\end{center}

Met \'Eireann, Ireland's meteorological service, operates the Dunsany weather station 7~km from the Woodland substation. At this weather station, rainfall measurements are taken. In addition, daily soil moisture deficit (SMD) measurements are taken for soil samples with different drainage characteristics (classed as well, moderately or poorly drained soils). This measures the amount of rain needed to bring the soil moisture content back to capacity. When the SMD is at a minimum measured value of $-10$~mm, the soil is considered saturated. The measured daily rainfall and SMD in a moderately drained soil sample for the 5 events are shown in Figure~\ref{RAINFALL}. Of all of the periods of interest, August 2015 had the most rainfall immediately before a geomagnetic storm, with 60.2~mm of rain falling in the three days prior to the event. On average, Ireland experiences between 2 and 2.7~mm of rainfall per day, and the heavy rainfall saturated the soil samples at Dunsany on 20 August 2015. The event with the next largest amount of rainfall was the October 2015 event, with 8.6~mm of rainfall in the three days before this event. 

\begin{center}
\begin{figure*}
\noindent\includegraphics[width=30pc]{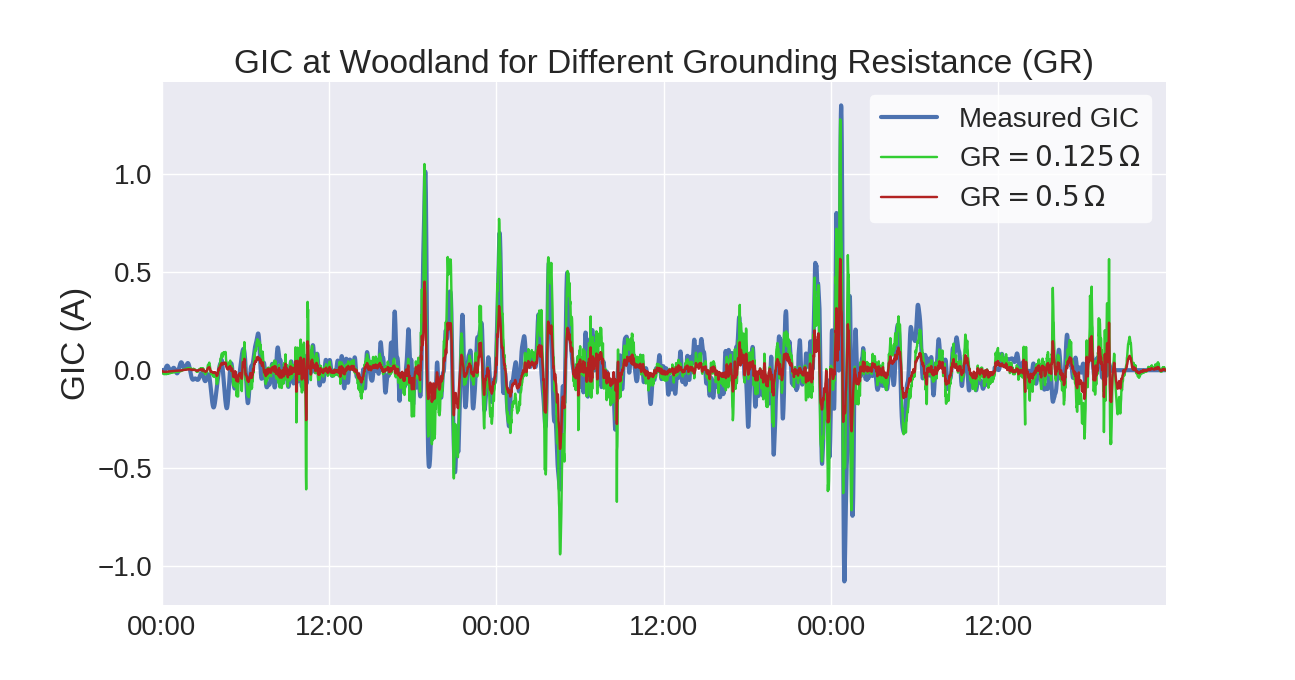}
\caption{Measured and calculated GIC at the Woodland transformer for different grounding resistances (GR) for the August 2015 geomagnetic storm. Changing this value from 0.5 to 0.125~$\Omega$ increases the amplitude of the calculated GICs to better match the measured GICs.}
\label{DIFF_GR}
\end{figure*}
\end{center}

It is possible that heavy rainfall prior to the August 2015 event had a measurable effect on the galvanic connection between the transformer grounds and the Earth. By reducing the grounding resistance at Woodland from 0.5~$\Omega$ to 0.125~$\Omega$ in the power network model, the calculated GICs better match the larger peaks in the measured GIC timeseries for the August 2015 event. This can be seen in Figure~\ref{DIFF_GR}. Despite better fitting the larger peaks, the RMSD actually decreased from at 0.127 to 0.122 with the lower grounding resistance. This is due to the overestimation of GIC amplitudes towards the end of the event event. This may correspond to the ground draining by 21 August 2015 (Figure~\ref{RAINFALL}), increasing the substation ground resistance value.

While the above exercise suggests that rainfall may have had some influence on the magnitude of GICs in Woodland for this event, a number of uncertainties exist. The moderately drained soil sample at the Dunsany weather station was saturated at the start of the August event. The SMD of the soil at and around the Woodland substation is unknown however, as is how the grounding resistance value may change over time. For this simulation, only the grounding resistance at the Woodland substation was changed. If rainfall had an effect on the grounding resistance at Woodland, it is likely that it had an effect on other substations. This was not factored in to the simulations, as GIC measurements are limited only to Woodland. There may also have been changes in the makeup of the network which could account for the differences in the five different storm simulations, and the rainfall may simply be coincidental.  Whether or not the rainfall is the cause of the mismatch between the simulated and measured GICs in this paper, it is reasonable to assume that grounding resistances at substations do not remain constant with varying soil moisture.

\section{Discussion and Conclusion}

The entire HV power network operating in Ireland and Northern Ireland was modelled for studying GICs. This model includes 400, 275, 220 and 110~kV substations and transmission lines. Substations operating at 220~kV or higher had the correct number, type and DC winding resistances of transformers modelled. Measured grounding resistances were also used for 33 of the substations across Ireland, with the remainder having assigned grounding resistances of 1~$\Omega$. An open-source Python program (https://doi.org/10.5281/zenodo.1252432) was written to use all of the available power network information and construct a network model consistent with the \citet{Lehtinen1985} method for calculating GICs. The resulting power network model for Ireland is an improvement on the simpler model seen in \citet{Blake2016}.

The improved power network model was subjected to uniform 1~V~km$^{-1}$ electric fields in a number of sensitivity tests. It was found that apart from a single 400~kV substation in the West of Ireland (Moneypoint), no substation had calculated GICs greater than 40~A. These are lower GIC amplitudes than seen for the same test in other European grids. This is likely due to the network's small size and the ground resistance assumption of 1~$\Omega$.

This 1~$\Omega$ assumption for substation grounding resistance was then investigated using uniform electric fields. It was found that different substations were affected to differing degrees by the grounding resistance assumption. For some substations, calculated GIC amplitudes were barely affected when the grounding resistance assumption was changed. In other substations, the grounding resistance was an important factor for calculated GIC amplitudes. Interestingly, this was found to be true even for some substations that had known, fixed grounding resistance values. One such substation, Moneypoint, had calculated GIC that changed by only 0.7~A when the grounding resistances were changed in the rest of the network. In contrast, the calculated GIC at Rafeen, (another substation with a known grounding resistance) changed by 30~A with the grounding resistance assumption. The calculated GIC at Rafeen could be said to depend on the grounding resistance of the rest of the network, whereas this is not the case for Moneypoint. 

It is not immediately clear why this dependence on grounding resistance exists for some substations in the network but not for others. No simple relation was found between the variability in GIC amplitudes (middle row, Figure~\ref{GROUND_GIC}) and the characteristics of the substation elements or connections to the rest of the network. By finding this relation, or alternatively by performing the ground resistance simulation described in Setion~\ref{GROUND_SECTION_REF} for a network, one can identify substations for which the ground resistivity is a particularly important factor for GIC generation. This could help prioritise which substations should have empirical ground resistances measured in the future.

It was also found that omitting lower-voltage elements of the power network model (in this case 110~kV substations and lines) led to general overestimation of GIC amplitudes in the network model. Those substations which were most affected by including the lower voltage elements were themselves connected to lower voltage substations and lines.

A comparison between the power network model in this paper and the model that appears in \citet{Blake2016} can be made, allowing us to reassess the vulnerability of individual substations. In \citet{Blake2016}, Moneypoint was assigned a grounding resistance of 0.1~$\Omega$, the same assumed value for every substation in that model. In the new network model, Moneypoint now has a grounding resistance value of 0.25~$\Omega$, the lowest in the network. This gives a computed GIC of 114~A for an eastward 1~V~km$^{-1}$ test electric field, a large increase compared to 41~A reported for Moneypoint in \citet{Blake2016}. As Moneypoint is Ireland's largest electricity generating substation, these larger GIC estimates are noteworthy.

The 275/220~kV Louth substation is an example of another substation which has to be reassessed with regards to GIC estimates. Previously, modest GICs were calculated for the Louth substation, with a peak of 9~A for a northward-oriented 1~V~km$^{-1}$ test electric field. With the latest network model, the GIC calculated at the Louth substation is now 38~A, second only to Moneypoint. Louth is treated differently in a number of ways in the new network model. As the Louth substation is now known to operate with three different voltages, it is modelled as two separate but spatially close substations (275/220 and 220/110~kV), with a total of seven transformers. These transformers have a lower average resistance than the single transformer used in the previous network model (0.17~$\Omega$ versus 0.5~$\Omega$), which may have contributed to the larger GICs, despite the larger ground resistance value of 1~$\Omega$ used in the new model. Another difference is the number of connections to the Louth substation. With the addition of the 110~kV substations and lines, Louth now has 10 transmission line connections to other substations, whereas before it had only four. For an eastward directed field, the addition of the 110~kV elements in the network reduces the calculated GIC at Louth significantly (see Figure~\ref{VOLTAGES}), as is the general trend in the network. For the case of a northward directed field however, the lower voltage elements slightly increase the calculated GIC at Louth. In terms of GICs, Louth is an important substation as it is the only connection between 275 and 220~kV elements, and sees the second largest GICs in the network. Along with Moneypoint, Louth should be prioritised in future studies on the Irish power network.

The updated model of the Irish power network was used to simulate GICs in the 400/220~kV Woodland substation for five geomagnetic storm events that occurred between 2015 and 2016. The measured GIC at Woodland was replicated with correlation coefficients ranging from 0.43 to 0.68 for the different events. GICs simulated for the August 2015 event underestimated measured GICs at the Woodland substation by a factor of 2.5, despite having the highest correlation coefficient of the events. In the three days prior to the August 2015 event, 60~mm of rain fell near to this substation, saturating the ground. This heavy rainfall may have reduced the substation grounding resistance. A grounding resistance change from 0.5~$\Omega$ to approximately 0.125~$\Omega$ at Woodland would account for much of the discrepancy between simulated and measured GIC peaks for this event. This event indicates that terrestrial weather may be an additional factor which can be taken into account for more complete GIC simulations.

While the power network model has been greatly improved by incorporating as much information as was available, assumptions were still made in order to make the model complete. Further information is required to make comprehensive GIC calculations in the future. This includes details on the transformer resistances in the 110~kV substations, and realtime information on the makeup of the power network (to account for elements being powered on and off due to maintenance). Substation grounding information is only known for the 33 of the 274 substations, and as demonstrated, the assumed grounding resistance value can greatly affect GIC calculations at individual substations. If (or when) this information becomes available in the future, it may be the case that our understanding of the vulnerability of different substations will change, as our understanding of the Louth and Moneypoint substations has changed with the latest iteration of the network model.

In addition, the uniform 400~$\Omega$m Earth model used for the geomagnetic storm event simulations is a significant simplification. As shown in this study, this model was sufficient for replicating measured GIC values in the Woodland substation. However, a uniform Earth model neglects both spatial variations in the surface electric field due to conductive variation, as well as the sea or coastal effect in areas close to shorelines. These can each affect GIC values at substations, and future GIC studies will take advantage of a full 3D Earth model for Ireland from long-period MT measurements. This future model is currently being developed as part of the Space Weather Electromagnetic Database for Ireland (SWEMDI) project, funded by the Geological Survey of Ireland.



\newpage

%
%
%
%
%
%
%

\begin{acknowledgments}
The results presented in this paper rely on data collected at magnetic observatories. We thank the national institutes that support them and INTERMAGNET for promoting high standards of magnetic observatory practice (www.intermagnet.org). We also acknowledge Armagh Observatory for hosting a magnetometer which contributed to this work. Historical rainfall and soil moisture deficit data for the Dunsany weather station was taken from Met \'Eireann's website (www.met.ie). Calculated and measured GICs for the events in this paper are given as supporting information. This research was funded by the Irish Research Council's Enterprise Partnership Scheme between Trinity College Dublin and EirGrid Plc. We thank the Irish Research Council for support for Se\'{a}n Blake and Joan Campany\`{a}.  This research benefited from a Royal Society/Royal Irish Academy International Exchange grant (IE150685). Ciar\'an Beggan, Gemma Richardson and Alan Thomson were supported by Natural Environment Research Council grant NE/P017231/1 'Space Weather Impacts on Ground-based Systems (SWIGS)'.

\end{acknowledgments}

%
%
%
%
%
%
%
%
%

\newpage

%
%

\end{article}





\end{document}